\documentclass[prl,twocolumn,superscriptaddress,showpacs,floatfix]{revtex4}
\usepackage{graphicx}

\newcommand{\be}{\begin{equation}}
\newcommand{\ee}{  \end{equation}}
\newcommand{\ba}{\begin{eqnarray}}
\newcommand{\ea}{  \end{eqnarray}}
\newcommand{\ve}{\varepsilon}

\begin{document}

\title{Distribution of Partial Neutron Widths for Nuclei Close to a
Maximum of the Neutron Strength Function}

\author{Hans A. Weidenm{\"u}ller}
\affiliation{Max--Planck--Institut f{\"u}r
Kernphysik \\ Heidelberg, Germany}

\begin{abstract}
For nuclei near a maximum of the neutron strength function, the
secular dependence on energy $E$ of $s$--wave partial neutron widths
differs from the canonical form $\sqrt{E}$. We derive the universal
form of that dependence and show that it is expected to significantly
influence the analysis of neutron resonance data.
\end{abstract}


\maketitle

{\it Purpose.} The Porter--Thomas distribution~\cite{Por56} is one of
the key predictions of Random--Matrix Theory (RMT). The reduced
partial neutron widths (simply ``neutron widths'' in the sequel) of
compound--nucleus (CN) resonances are predicted to follow a $\chi^2$
distribution with $\nu = 1$ degrees of freedom. That prediction was
recently tested with unprecedented accuracy~\cite{Koe10}. The authors
scattered slow neutrons on several Pt isotopes, thereby measuring
sequences of CN resonances over energy intervals of up to $20$ keV
length, and obtained sets of widths for $s$--wave neutrons containing
up to $450$ data points. Reduced neutron widths were obtained by
rescaling the measured widths by the factor $f^2(E) = \sqrt{E}$ with
$E$ taken at resonance energy. That factor is supposed to take account
of the secular variation with energy of the widths for $s$--wave
neutrons.  Using a cutoff procedure to minimize $p$--wave background
and a maximum--likelihood analysis, the authors concluded that the
validity of the Porter--Thomas distribution must be rejected with a
statistical significance of at least $99.997$ per
cent~\cite{Koe10}. That result calls into question earlier successful
tests of RMT in nuclei (for a review, see Ref.~\cite{Wei09}). More
generally, it questions whether at excitation energies of several MeV
nuclei are correctly described as basically chaotic systems, a view
widely held so far. Not surprisingly, the result announced by Koehler
{\it et al.}  has found wide attention~\cite{Rei10}.

For the isotopes of Pt investigated in Ref.~\cite{Koe10}, the neutron
strength function (the ratio of the average neutron partial width and
the mean resonance spacing) is strongly enhanced. The enhancement
facilitates the separation of $s$--wave resonances from the
contamination of small $p$--wave resonances. To the best of our
knowledge, the theoretical implications of that enhancement for the
secular dependence of neutron widths on energy, i.e., for the function
$f^2(E)$ introduced above, have never been investigated. In this
Letter we show that the enhancement of the strength function implies
that $f^2(E)$ carries an additional energy dependence on top of the
$\sqrt{E}$--dependence mentioned above. That dependence has never been
taken into account in any analysis of data on neutron widths. It is
not clear whether the conclusions drawn in Ref.~\cite{Koe10} will be
upheld when that dependence is accounted for.

Maxima of the neutron strength function are due to a resonance close
to threshold (more precisely: to a virtual state as defined below) or
to a weakly bound state of the nuclear single--particle potential for
$s$--wave neutrons. Such maxima occur systematically with increasing
mass number (i.e., increasing radius of the nuclear single--particle
potential) when a bound $s$--wave state is being formed at
threshold. (For the Pt isotopes that is the $4s$--state in
spectroscopic notation). The enhancement of neutron widths due to the
nascent bound single--particle state is strongest for CN resonances
close to neutron threshold and, being a threshold effect, dies out
with increasing separation of the CN resonances from neutron
threshold. Such weakening of the enhancement causes the additional
energy dependence of $f^2(E)$ and forms the topic of the present
paper.

To make the paper self--contained we begin with a brief account of
single--channel CN scattering theory. We show that $f(E)$ is
determined by the projection of the neutron single--particle
scattering wave function onto the nuclear volume. We calculate that
projected part of the wave function for an attractive square--well
potential and determine $f(E)$ for both, a single--particle resonance
close to threshold and a weakly bound state. We show that the
enhancement is a generic feature and occurs likewise in more realistic
nuclear single--particle potentials that differ from a square
well. Whenever the $s$--wave strength function shows significant
enhancement, $f^2(E)$ differs significantly from $\sqrt{E}$ even over
energy intervals as small as $20$ keV (the interval used in
Ref.~\cite{Koe10}).

{\it Single--Channel CN Scattering.} We deal with a single channel
(the $s$--wave neutron channel). The energy $E$ is positive and $E =
0$ denotes the threshold. In the shell--model approach to nuclear
reactions~\cite{Mah69}, the channel wave function $\chi_E$ is the
antisymmetrized product of the wave function of the target nucleus in
its ground state and the neutron single--particle scattering wave
function $\Psi_E(r)$. The latter depends only on the radial coordinate
$r$ and is chosen real. The numerous many--body resonances typical for
CN scattering are due to $N \gg 1$ orthonormal quasibound states $|
\mu \rangle$ where $\mu = 1, 2, \ldots, N$. These interact with each
other through the $N$--dimensional real and symmetric Hamiltonian
matrix $\langle \mu | H | \nu \rangle = H_{\mu \nu}$ and are coupled
to the neutron channel by real Hamiltonian matrix elements
\be
W_\mu(E) = \langle \chi_E | H | \mu \rangle \ .
\label{0}
\ee
The unitary scattering amplitude $S(E)$ that takes into account the
presence of the $N$ CN resonances has the form~\cite{Mah69,Mit10}
\be
S(E) = \exp [ 2 i \delta ] \bigg( 1 - 2 i \pi \sum_{\mu, \nu = 1}^N
W_\mu(E) D^{-1}_{\mu \nu}(E) W_\nu(E) \bigg) \ .
\label{1}
\ee
Here $\delta$ is the $s$--wave potential scattering phase shift, and
\be
D_{\mu \nu}(E) = E \delta_{\mu \nu} - H_{\mu \nu} - F_{\mu \nu}(E) +
i \pi W_\mu(E) W_\nu(E)
\label{2}
\ee
where
\be
F_{\mu \nu}(E) = \int_0^\infty {\rm d} E' \ \frac{\cal P}{E - E'}
W_\mu(E') W_\nu(E')
\label{3}
\ee
and where ${\cal P}$ denotes the principal--value integral. The last
term in Eq.~(\ref{2}) is proportional to the width matrix and descibes
the instability of the CN resonances due to their coupling to the
channel. The matrix $F_{\mu \nu}(E)$ in Eq.~(\ref{3}) accounts for the
shift of the resonances due to that coupling. Eqs.~(\ref{1}) to
(\ref{3}) provide a very general framework for $s$--wave neutron
scattering in the presence of CN resonances. The assumption that the
CN resonances are described by RMT is implemented by postulating that
the matrix $H_{\mu \nu}$ is a member of the Gaussian Orthogonal
Ensemble (GOE) of random matrices.

Near neutron threshold the CN resonances are isolated, and we use the
diagonal representation $H_{\mu \nu} = \sum_\rho {\cal O}_{\rho \mu}
E_\rho {\cal O}_{\rho \nu}$ where ${\cal O}_{\mu \nu}$ is orthogonal
and where $E_\rho$ are the eigenvalues of $H_{\mu \nu}$. We define
$\tilde{W}_{\mu}(E) = \sum_\nu {\cal O}_{\mu \nu} W_{\nu}$ and
$\tilde{F}$ by Eq.~(\ref{3}) with all $W$s replaced by $\tilde{W}$s.
For isolated resonances the coupling between the eigenvalues $E_\rho$
is negligible (both $\tilde{F}_{\mu \nu}$ and $\tilde{W}_\mu
\tilde{W}_\nu$ are taken to be diagonal), and $S(E)$ takes the form
\be
S(E) = \exp [ 2 i \delta ] \bigg( 1 - 2 i \pi \sum_{\mu}
\tilde{W}_\mu(E) (E - {\cal E}_\mu)^{-1} \tilde{W}_\mu(E) \bigg)
\label{6}
\ee
where the complex resonance energies ${\cal E}_\mu$ are given by
\be
{\cal E}_\mu = E_\mu + \tilde{F}_{\mu \mu} - i \pi \tilde{W}^2_\mu \ .
\label{6a}
\ee
The neutron partial width amplitude of the resonance located at ${\cal
E}_\mu$ is given by $\sqrt{2 \pi} \tilde{W}_\mu(E) = \sqrt{2 \pi}
\sum_\nu {\cal O}_{\mu \nu} W_\mu(E)$. In order to remove any secular
energy dependence from the matrix elements $W_\mu(E)$ we write
\be
W_\mu(E) = f(E) V_\mu \ .
\label{7}
\ee
By construction, the amplitudes $V_\mu$ do not carry any secular
dependence on energy $E$. GOE predicts that in the limit of infinite
matrix dimension, the elements ${\cal O}_{\mu \nu}$ and, therefore,
the amplitudes $\sum_\nu {\cal O}_{\mu \nu} V_\nu$ are
Gaussian--distributed random variables. As a consequence, the reduced
neutron widths $2 \pi \tilde{W}^2_\mu(E) / f^2(E)$ are predicted to
follow the Porter--Thomas distribution. To test that prediction, we
must determine the function $f(E)$ for nuclei in the vicinity of a
single--particle $s$--wave resonance close to threshold and for a
weakly bound single--particle $s$--wave state.

{\it Poles of the Single--Particle Scattering Amplitude.}
Eq.~(\ref{0}) shows that the entire energy dependence of $W_\mu(E)$ is
due to $\chi_E$. We recall that $\chi_E$ is the antisymmetrized
product of the ground--state wave function of the target nucleus and
the real $s$--wave scattering wave function $\Psi_E(r)$. For the
matrix element $W_\mu(E)$, only the projection of $\Psi_E$ onto the
nuclear volume is relevant. Thus, the energy dependence of $W_\mu(E)$
is determined by the enery dependence of $\Psi_E(r)$ for $r \leq R$
where $R$ is the nuclear radius.

The scale in energy over which the radial dependence of a
single--particle scattering wave function changes, is given by the
typical distance between bound $s$--wave single--particle states. In a
heavy nucleus, that distance is about $10$ MeV and, thus, very large
compared to the typical energy scale over which resonance data are
taken. (In Ref.~\cite{Koe10}, that scale was $20$ keV). Therefore, the
radial form of the neutron $s$--wave scattering function in the
nuclear volume $r \leq R$ can safely be taken as independent of
energy, and we focus attention on the energy--dependent amplitude
$f(E)$ of that function. In the vicinity of the threshold energy ($E =
0$), $f(E)$ is strongly enhanced whenever a single--particle $s$--wave
resonance or a bound $s$--wave single--particle state occurs close to
threshold. Both resonance and bound state manifest themselves as poles
of the unitary single--particle potential scattering amplitude $s(E) =
\exp [ 2 i \delta ]$ in Eq.~(\ref{1}). To understand qualitatively
what happens we recall some properties of the poles of $s(E)$. The
potential scattering wave function $\Psi_E(r)$ at energy $E$ depends
asymptotically on wave number $k$ where $\hbar^2 k^2 / (2 m) = E$, and
it is useful to consider the distribution of poles of the scattering
amplitude $s(E)$ in the complex $k$--plane (rather than the complex
energy plane). We accordingly replace $s(E)$ by $s(k)$.

For a square--well potential, poles of $s(k)$ have been studied in
detail~\cite{Nus59,Mah69}. Poles of $s(k)$ occur either pairwise or as
single poles. Pairs of poles lie below the real $k$--axis and occur
symmetrically to the imaginary $k$--axis. Such pairs are, thus,
located at $k_0$ and at $- k^*_0$ with $\Im (k_0) \leq - i/a$ (where
$a$ is the radius of the square--well potential) and $\Re (k_0) >
0$. Single poles lie on the imaginary $k$--axis. Poles on the positive
imaginary $k$--axis correspond to bound states while poles on the
negative imaginary $k$--axis are referred to as virtual states. We
visualize the motion of the poles in the complex $k$--plane as the
depth $V_0$ of the square--well potential is increased. For a very
shallow potential, there are no bound states and no poles on the
imaginary $k$--axis. All poles lie far below the real $k$--axis, and
significant resonant behavior is absent. As $V_0$ is increased, the
first pair of poles approaches the negative imaginary axis from
opposite sides. The two poles coalesce at $k = - i/a$. Then one pole
moves up and the other pole moves down the imaginary $k$--axis.
Significant resonance behavior of $s(k)$ is caused only by the
upward--moving pole, first as a virtual state and later as a weakly
bound state. (For $a = 6$ fm, a realistic value for the radius of a
heavy nucleus, the point $k = - i/a$ has a distance in energy of about
$0.5$ MeV from threshold). As $V_0$ is increased further, resonance
enhancement subsides. The pattern repeats itself as the next pair of
poles approaches the point $k = - i/a$ and the $2s$--state is pulled
into the potential, and so on.

{\it Enhancement factor for the square--well potential}. We work out
the resulting enhancement of the amplitude $f(E)$ for the $4s$--state
although the analysis and result are exactly the same for any
$s$--state near threshold. The real scattering function $\Psi_E(r)$ is
normalized to a delta function in energy. An elementary calculation
shows that for $r < a$ we have $\Psi_E(r) = f(E) \sin(\kappa r) / r$
where $\hbar^2 \kappa^2 / (2 m) = E + V_0$ and where
\be f(E) = \sqrt{\frac{m}{\pi k \hbar^2}} \frac{2 (k a)}{\sqrt{(k a)^2
\sin^2 (k a) + (\kappa a)^2 \cos^2 (\kappa a)}} \ .
\label{8}
\ee
As remarked above, the function $\sin (\kappa r)$ changes very slowly
with energy, and attention is focused on $f(E)$.

A weakly bound $4s$--state with energy binding $E_0 < 0$ exists
if the condition $[(k a) / (\kappa a)] \tan (\kappa a) = - 1$ is met
for $(\kappa a) = (7 \pi / 2) + \ve$ and $\ve \ll 1$. Then $E_0 = - (7
\pi / 2)^2 (\hbar^2 \ve^2) / (2 m a^2)$. Expanding $(\kappa a)$, $\sin
(\kappa a)$ and $\cos (\kappa a)$ in Taylor series around $\kappa a =
7 \pi / 2$ and keeping only lowest--order terms, we find
\be
f^2(E) \propto \frac{\sqrt{E}}{E + |E_0|} \ .
\label{9}
\ee
We have suppressed energy--independent factors. The factor $\sqrt{E}$
is universal for $s$--wave scattering near threshold. The factor $(E +
|E_0|)^{-1}$ describes the enhancement due to the weakly bound
single--particle $s$--wave state. For the virtual state the condition
reads $[(k a) / (\kappa a)] \tan (\kappa a) = + 1$. That yields for
$f^2(E)$ the same form as in Eq.~(\ref{9}) except that now $E_0$ is
the energy associated with the virtual state on the negative imaginary
$k$--axis.

We have expanded $(\kappa a)$ in powers of $(k a)$ around the location
of the pole of $s(E)$, and we have kept only the term of zeroth
order. The term of next order is $(k a)^2 / (7 \pi)$. For $0 \leq E
\leq 100$ keV and $a = 6$ fm, that term is not larger than $0.01$, and
we expect the pole approximation in Eq.~(\ref{9}) to be excellent.

{\it Discussion.} Although derived specifically for a square--well
potential, the factor given in Eq.~(\ref{9}) is universal and
describes for $r \leq R$ the enhancement of the square of the
$s$--wave single--particle scattering wave function also for other,
more realistic forms of the single--particle potential. Indeed, the
pattern of movement of the poles of $s(k)$ in the complex $k$--plane
versus depth of the potential is the same for all potentials that lack
a barrier. Pairs of $s$--wave poles occur some distance below the real
$k$--axis and symmetrically to the imaginary axis and cannot give rise
to significant resonance enhancement. Therefore, the value of $k$
where such pairs coalesce on the negative imaginary $k$--axis,
although different from that of the square--well potential, amounts to
at least several $100$ keV.  Significant resonance behavior is again
due to the pole of $s(k)$ that moves up on the imaginary $k$--axis
from the point of coalescence, first as a virtual and then as a weakly
bound state. Potential scattering theory~\cite{Mah69} shows that at a
pole of $s(k)$, $\Psi_E(r)$ is singular. For $k$ positive and close to
a pole on the imaginary $k$--axis, $f^2(E)$ has the form of
Eq.~(\ref{9}).

The universality of our result is displayed by the fact that $f^2(E)$
in Eq.~(\ref{9}) depends only on the energy of the virtual or weakly
bound state and is independent of any detailed properties of the
potential. (The value of $E_0$ in Eq.~(\ref{9}) does, of course,
depend on the potential). The denominator in Eq.~(\ref{9}) is
obviously a special case of the universal Lorentzian factor $[(E -
E_0)^2 + (1/4) \Gamma^2]^{-1/2}$ describing resonance enhancement and
applies when the resonance is located below threshold so that $E_0$ is
negative and $\Gamma$ vanishes. We also note that the enhancement
factor in Eq.~(\ref{9}) is similar to that due to a doorway state.

We conclude that for a virtual or a weakly bound neutron $s$--wave
state, the partial neutron widths carry the universal enhancement
factor given in Eq.~(\ref{9}). The enhancement factor affects both,
the neutron strength function and the secular energy dependence of the
widths of neutron $s$--wave resonances. To see what happens to the
neutron strength function, we keep $E$ (or $k$) fixed, positive, and
slightly above threshold, and we increase mass number $A$, thereby
increasing the radius of the potential. A pair of $s$--wave neutron
resonances approaches the imaginary $k$--axis, eventually giving rise
to a virtual and, later, to a bound state. For the virtual state, the
energy $|E_0|$ decreases monotonically toward zero and then, as the
virtual state turns into a bound state, increases monotonically from
zero. The result is first a steady increase of the enhancement
factor~(\ref{9}) and then, after the virtual state becomes a bound
state, a decrease of that factor. Taken together, that causes the
characteristic maximum of the strength function. Conversely, when
neutron widths are measured near threshold for a sequence of $s$--wave
resonances at some fixed value of $A$ where the strength function is
enhanced and, therefore, the enhancement factor in Eq.~(\ref{9}) is
operative, the resulting secular energy dependence of the neutron
widths goes beyond the standard $\sqrt{E}$--dependence.

Given the universality of the enhancement factor in Eq.~(\ref{9}), we
ask how that factor is expected to affect the analysis of neutron
resonance data. We take the work of Ref.~\cite{Koe10} as an example.
We assume that the distance of $|E_0|$ from threshold is large
compared to the average resonance spacing. We recall that the data
typically range over an energy interval of $20$ keV. To predict the
influence of the enhancement factor in Eq.~(\ref{9}) on the data, one
would have to determine $E_0$ with an accuracy of about $100$
keV. That requirement precludes a theoretical prediction using the
neutron single--particle potential in heavy nuclei. That potential is
not known to such an accuracy. Therefore, we use another estimate. The
mass dependence of single--particle levels is governed by the factor
$A^{- 2/3}$. For a potential depth of several ten MeV and $A \approx
200$, a weakly bound single--particle level changes by about $150$ keV
when $A$ changes by one unit. Therefore, we expect that for nuclei
near the maximum of the neutron strength function, $|E_0|$ is of the
order of $100$ keV. That is also consistent with the fact that for the
square--well potential, the point of coalscence has a distance of
about $500$ keV from threshold.  Taking in Eq.~(\ref{9}) $|E_0| = 50$
keV ($10$ keV) as examples, we see that the resonance enhancement
factor (the denominator of expression~(\ref{9})) changes by a factor
$1.4$ (a factor $3$, respectively) over an interval of $20$ keV
starting from threshold. Thus, the resonance enhancement in
Eq.~(\ref{9}) may significantly influence the analysis of data on
neutron widths.

In such an analysis, a first estimate of $E_0$ can be obtained from
Eq.~(\ref{9}) and the measured enhancement of the strength function.
One may then consider $E_0$ and the average value of the reduced
neutron partial widths as free parameters, and search for a best fit
to a $\chi^2$ distribution with $\nu$ degrees of freedom for the
reduced partial widths. Alternatively, it is possible to determine the
enhancement factor directly, i.e., without using the pole
approximation of Eq.~(\ref{9}). That can be done by solving
numerically the radial Schr{\"o}dinger equation for $s$--wave neutrons
for a potential that is realistic in form and that possesses a virtual
or a bound state close to threshold. The factor $f(E)$ is then
determined by the delta--function normalization condition for the
solution of the radial equation. We stress that the single--particle
nuclear potential to be used is not the optical--model potential
(which describes the {\it average} neutron scattering amplitude and,
thus, incorporates the effect of the CN resonances) nor is it the real
part of the optical--model potential (because the imaginary part
contributes to the real part via a dispersion relation). Rather, it is
the pure shell--model potential for neutrons. Conclusions about the
failure of RMT can be drawn only if such an approach yields a value
for $\nu$ that is significantly different from unity. Conversely, if
agreement is obtained for $\nu \approx 1$, that should make it
possible to determine the energy $E_0$ fairly precisely (with an error
of perhaps not more than $100$ keV) and, from there, the shell--model
potential for neutrons in heavy nuclei with great accuracy.

We mention in passing that the resonance enhancement in Eq.~(\ref{9})
also affects the shift matrix $F_{\mu \nu}(E)$ in Eq.~(\ref{3}). With
the same approximations as used above, the level shift $\tilde{F}_{\mu
\mu}$ in Eq.~(\ref{6a}) is given by the product of the reduced partial
width for the resonance at ${\cal E}_\mu$ and the factor
$\int_0^\infty {\rm d} E' \ {\cal P} f^2(E') / (E_\mu - E')$. The
enhancement of the shift due to the pole of $f^2(E)$ in Eq.~(\ref{9})
may be considerable and have important implications for the analysis
of spectral fluctuations.

So far we have considered the case where $|E_0|$ is much larger than
the mean spacing of the resonances. That is probably the typical
case. If that condition fails (i.e., if $|E_0| \leq 5$ keV or so), it
is not justified to consider the neutron widths as energy--independent
constants. Then it is essential to display the full energy dependence
of $S(E)$ explicitly. We use Eqs.~(\ref{7}) and (\ref{9}) and rewrite
the last term in Eq.~(\ref{6}) as
\be
- 2 i \pi \sum_{\mu} \sqrt{E} \tilde{V}_\mu [(E + | E_0|)(E - E_\mu
- \tilde{F}_{\mu \mu}) + i \pi \sqrt{E} \tilde{V}^2_\mu]^{-1}
\tilde{V}_\mu \ .
\label{10}
\ee
Here $\tilde{V}_\mu = \sum_\nu {\cal O}_{\mu \nu} V_\nu$.
Eq.~(\ref{10}) is qualitatively different from Eq.~(\ref{6}) with
energy--independent widths. The formulas of $R$--matrix theory
commonly used for the analysis of neutron cross--section data bear a
close analogy to the latter equation~\cite{Mah69}. Eq.~(\ref{10})
shows that these formulas cannot be used when $E_0$\ is very close to
threshold.

In summary, we have shown that a substantial enhancement of the
neutron strength function necessarily implies a significant energy
dependence of the neutron partial widths for resonances in the
vicinity of neutron threshold. We have derived the universal form of
that energy dependence. Conclusions about the validity of RMT
predictions can reliably be drawn only when that dependence is taken
into account in the analysis of neutron resonance data.

{\it Acknowledgments.} I am grateful to E. Reich for drawing my
attention to the work of Koehler {\it et al.} and to O. Bohigas,
G. E. Mitchell, P. E. Koehler, and F. Becvar for stimulating
correspondence.

\end{document}